\documentclass[times]{aastex631}
\usepackage{amssymb}
\usepackage{amsmath}
\usepackage{gensymb}
\usepackage{graphicx}
\usepackage{xcolor}
\usepackage{natbib}

\newcommand{\beq}{\begin{equation}}
\newcommand{\eeq}{\end{equation}}

\newcommand{\Ms}{\textrm{M}_*}
\newcommand{\Msun}{\textrm{M}_\odot}
\newcommand{\kmps}{km~s$^{-1}$}
\newcommand{\MHI}{\rm{M_{H{\textsc i}}}}

\newcommand{\MB}{{\rm M_B}}

\newcommand{\htwo}{${\rm H_2}$}

\newcommand{\hi}{H{\sc i}}

\newcommand{\hii}{H{\sc i} 21\,cm}

\newcommand{\tdephi}{{\rm t_{dep,H{\textsc i}}}}

\newcommand{\fhi}{{\rm f_{\rm H{\textsc i}}}}

\graphicspath{{./}{figures/}}

\shorttitle{\hi\ properties at $z\approx1$}
\shortauthors{Chowdhury, Kanekar and Chengalur}

\begin{document}

	
	\title{Atomic Gas Scaling Relations of Star-forming Galaxies at $z\approx1$}	

	\correspondingauthor{Aditya Chowdhury}
	\email{chowdhury@ncra.tifr.res.in}
	
	\author{Aditya Chowdhury}
	\affil{National Centre for Radio Astrophysics, Tata Institute of Fundamental Research, Pune, India.}
	
	\author{Nissim Kanekar}
	\affil{National Centre for Radio Astrophysics, Tata Institute of Fundamental Research, Pune, India.}
	
	\author{Jayaram N. Chengalur}
	\affil{National Centre for Radio Astrophysics, Tata Institute of Fundamental Research, Pune, India.}

	
	
	\begin{abstract}
    We use the Giant Metrewave Radio Telescope (GMRT) Cold-H{\sc i}  AT $z\approx1$ (CAT$z1$) survey, a 510~hr H{\sc i} 21cm\ emission survey of galaxies at $z=0.74-1.45$, to report the first measurements of atomic hydrogen (H{\sc i}) scaling relations at $z\approx1$. We divide our sample of 11,419 blue star-forming galaxies at $z\approx1$ into three stellar mass ($\textrm{M}_*$) subsamples and obtain detections (at $\geq 4\sigma$ significance) of the stacked H{\sc i} 21cm\ emission signal from galaxies in all three subsamples. We fit a power-law relation to the measurements of the average H{\sc i}  mass ($\textrm{M}_{\rm H{\textsc i}}$) in the three stellar-mass subsamples to find that the slope of the $\textrm{M}_{\rm H{\textsc i}}-\textrm{M}_{*}$ relation at $z\approx1$ is consistent with that at $z\approx0$. However, we find that the $\textrm{M}_{\rm H{\textsc i}}-\textrm{M}_{*}$ relation has shifted downwards from $z\approx1$ to $z\approx0$, by a factor of $3.54\pm0.48$. Further, we find that the \hi\ depletion timescales (${\rm t_{dep,H{\textsc i}}}$) of galaxies in the three stellar-mass subsamples are systematically lower than those at $z\approx0$, by factors of $\approx2-4$. 
      We divide the sample galaxies into three specific star-formation rate (sSFR) subsamples, again obtaining $\geq 4\sigma$ detections of the stacked H{\sc i} 21cm\ emission signal in all three subsamples. We find that the relation between the ratio of H{\sc i}  mass to stellar mass and the sSFR evolves between $z\approx1$ and $z\approx0$. Unlike the efficiency of conversion of molecular gas to stars, which does not evolve significantly with redshift, we find that the efficiency with which H{\sc i}  is converted to stars is much higher for star-forming galaxies at $z\approx1$ than those at $z\approx0$.	\end{abstract}
	
	
\section{Introduction}

Measurements of the neutral atomic hydrogen (\hi) properties of galaxies as a function of their redshift, environment, and stellar properties are important to obtain a complete picture of galaxy evolution. In the local Universe, the \hi\ properties of galaxies are known to depend on their global stellar properties, e.g. the stellar mass ($\Ms$), the star-formation rate (SFR), etc.  \citep[see][for a review]{Saintonge22}. Such ``\hi\ scaling relations'' at $z \approx 0$ serve as critical benchmarks for numerical  simulations and semi-analytical models of galaxy formation and evolution \citep[e.g.][]{Lagos18,Diemer18,Dave19}. 

Unfortunately, the faintness of the \hii\ line has severely hindered the use of \hii\ emission studies to probe the \hi\ properties of galaxies at cosmological distances. Even very deep integrations with today's best radio telescopes \citep[e.g.][]{Jaffe13,Catinella15,Gogate20} have yielded detections of \hii\ emission from individual galaxies out to only $z\approx0.376$ \citep{Fernandez16}. Thus, until very recently, nothing was known about the \hi\ properties of high-$z$ galaxies and how the \hi\ properties depend on the stellar mass, the SFR, or other galaxy properties.

The above lack of information about \hi\ scaling relations at high redshifts has meant that simulations of galaxy evolution are not well constrained with regard to gas properties beyond the local Universe. Specifically, while a number of simulations broadly reproduce the \hi\ scaling relations at $z \approx 0$ \citep[e.g.][]{Lagos18,Diemer18,Dave19}, the predictions for the evolution of these relations differ significantly \citep[e.g.][]{Dave20}. Measurements of \hi\ scaling relations at $z\gtrsim1$, along with similar relations for the molecular component \citep[e.g.][]{Tacconi20}, would hence provide a crucial benchmark for simulations of galaxy evolution. Further, such \hi\ scaling relations at $z\approx1$ would be useful in estimating the individual \hi\ masses of galaxies at these redshifts, and the sensitivity of upcoming \hii\ surveys to both individual and stacked \hii\ emission from galaxies at high redshifts \citep[e.g.][]{Blyth16}.

The \hii\ stacking approach \citep{Zwaan00,Chengalur01}, in which the \hii\ emission signals from a large number of galaxies with accurate spectroscopic redshifts are co-added to measure the average \hi\ mass of a galaxy sample, can be used to overcome the intrinsic weakness of the \hii\ line \citep[e.g.][]{Lah07,Delhaize13,Rhee16,Kanekar16,Bera19,Sinigaglia22}. This  approach has been used to measure the global \hi\ properties of local Universe galaxies as a function of their global stellar properties \citep[e.g.][]{Fabello11,Brown15,Guo21}. The \hi\ scaling relations obtained from these stacking analyses have been shown to be consistent with those derived from individual \hii\ detections \citep[e.g.][]{Saintonge22}. It should thus be possible to use the \hii\ stacking approach to determine the \hi\ scaling relations at cosmological distances \citep[e.g.][]{Sinigaglia22}.

\hii\ stacking experiments with the Giant Metrewave Radio Telescope (GMRT) have recently been used to measure the average \hi\ properties of blue star-forming galaxies at $z\gtrsim1$ \citep{Chowdhury20,Chowdhury21}. These studies have shown that  star-forming galaxies at $z\approx1$ have large \hi\ masses but that the \hi\ reservoirs can sustain the high SFRs of the galaxies for a short period of only $1-2$~Gyr. More recently, \citet{Chowdhury22a} used the GMRT Cold-\hi\ AT $z\approx1$ \citep[GMRT-CAT$z1$;][]{Chowdhury22b} survey, a 510~hr GMRT \hii\ emission survey of the DEEP2 fields \citep{Newman13}, to find that the average \hi\ mass of star-forming galaxies declines steeply by a factor of $\approx3.2$ from $z\approx1.3$ to $z\approx1.0$, over a period of $\approx1$~Gyr. This is direct evidence that the the rate of accretion of gas from the circumgalactic medium (CGM) on to galaxies at $z\approx1$ was insufficient to replenish their \hi\ reservoirs, causing a decline in the star-formation activity of the Universe at $z\lesssim1$. Subsequently, \citet{Chowdhury22c} used the GMRT-CAT$z1$ measurements of the average \hi\ mass of galaxies at $z\approx1.0$ and $z\approx1.3$ to show that \hi\ dominates the baryonic content of high-$z$ galaxies.

In this \emph{Letter}, we use the GMRT-CAT$z1$ survey to report, for the first time, measurements of \hi\ scaling relations at $z\approx1$, at the end of the epoch of peak cosmic star-formation activity in the Universe. 


\section{Observations and Data Analysis}
\subsection{The GMRT-CAT$z1$ Survey}

The GMRT-CAT$z1$ survey \citep{Chowdhury22b} used $\approx$510~hrs with the upgraded GMRT $550-850$~MHz receivers to carry out a deep \hii\ emission survey of galaxies at $z=0.74-1.45$, in three sky fields covered by the DEEP2 Galaxy Survey \citep{Newman13}. The three DEEP2 fields covered by the CAT$z1$ survey contain seven sub-fields of size $\approx 52' \times 28'$, each of which was covered using a single GMRT pointing. The design, the observations, the data analysis, and the main sample of galaxies of the GMRT-CAT$z1$ survey are described in detail in \citet{Chowdhury22b}. We provide here a summary of the information directly relevant to this paper.
		
The observations for the GMRT-CAT$z1$ survey were obtained over three GMRT observing cycles. The data of each subfield from each observing cycle were analysed separately. { This was done to prevent systematic effects present in the data of one cycle (e.g. low-level RFI, deconvolution errors, etc), from affecting the quality of the data from the other cycles \citep[see][for a detailed discussion]{Chowdhury22b}.} The analysis resulted in $2-3$ spectral cubes for each of the seven DEEP2 fields. The cubes have channel widths of $48.8$~kHz, corresponding to a velocity resolution of $18$~\kmps$-25$~\kmps, over the redshift range $z=0.74-1.45$. The FWHM of the synthesized beams of the spectral cubes are $4\farcs0-7\farcs5$ over the frequency range $580-830$~MHz, corresponding to spatial resolutions in the range $29$~kpc$-63$~kpc\footnote{Throughout this work, we use a flat ``737'' Lambda-cold dark matter cosmology, with $\Omega_m=0.3$, $\Omega_\Lambda = 0.7$, and $H_0 = 70$~\kmps~Mpc$^{-1}$.} for galaxies at $z=0.74-1.45$. 

 The GMRT-CAT$z1$ survey covers the \hii\ line for 16,250 DEEP2 galaxies with accurate spectroscopic redshifts \citep[velocity errors~$\lesssim 62$~\kmps;][]{Newman13} at $z=0.74-1.45$. We excluded (i)~red galaxies, identified using a cut in the $\rm (U-B)$ vs $\MB$ colour-magnitude diagram \citep{Willmer06,Chowdhury22b}, (ii)~radio-bright AGNs, detected in our radio-continuum images at $>4\sigma$ significance with rest-frame 1.4~GHz luminosities $\textrm{L}_{1.4 \textrm{GHz}}\ge2\times10^{23}$~W~Hz$^{-1}$ \citep{Condon02}, (iii)~galaxies with stellar masses $\Ms<10^9~\Msun$, and (iv)~galaxies whose \hii\ subcubes were affected by discernible systematic effects \citep{Chowdhury22b}. This yielded a total of 11,419 blue star-forming galaxies with $\Ms\geq10^9~\Msun$ at $z=0.74-1.45$, the main sample of the GMRT-CAT$z1$ survey. The  survey provides a total of 28,993 \hii\ subcubes for the 11,419 galaxies. The subcube of each galaxy covers a region of $\pm500$~kpc around the galaxy location, with a uniform spatial resolution of 90~kpc, and a velocity range of $\pm1500$~\kmps\ around its redshifted \hii\ frequency, with a channel width of 30~\kmps. { The median spectral RMS noise on the 28,993 \hii\ subcubes is $297 \ \mu$Jy per 30~\kmps\ velocity channel, at a spatial resolution of 90~kpc.}

{ We note that the average \hii\ emission signal from the sample of 11,419 galaxies is consistent with being unresolved at a spatial resolution of 90~kpc \citep{Chowdhury22b}. Further, the compact resolution of 90~kpc ensures that the average \hii\ emission signal does not include a significant contribution from companion galaxies in the vicinity of the target galaxies \citep{Chowdhury22b}.}
 
The stellar masses of the individual DEEP2 galaxies were obtained using a relation between the stellar mass\footnote{All stellar masses and SFRs in this work assume a Chabrier initial mass function (IMF). Estimates in the literature that assume a Salpeter IMF were converted a Chabrier IMF by subtracting 0.2~dex \citep[e.g.][]{Madau14}.} and the absolute rest-frame B-band magnitude ($\MB$), the rest-frame (U$-$B) colour, and the rest-frame (B$-$V) colour \citep{Weiner09}. The relation was calibrated using a subset of the DEEP2 galaxies with K-band estimates of the stellar masses \citep{Weiner09}. The SFRs of the individual galaxies were inferred from their $\MB$ values and rest-frame (U$-$B) colours, via the SFR calibration of \citet{Mostek12}; these authors used the SFRs of galaxies in the Extended Groth Strip (obtained via spectral-energy distribution (SED) fits to the rest-frame ultraviolet, optical, and near-IR photometry; \citealp{Salim09}) to derive the SFR calibration for the DEEP2 galaxies. \citet{Mostek12} found that the scatter between the SED SFRs of \citet{Salim09} and the SFRs obtained via the calibration based on the $\MB$ and (U$-$B) values is $\approx0.2$~dex\footnote{We note that we used the SFR calibration of \citet{Mostek12} that relates the SFR of a galaxy to its $\MB$, (U$-$B), and (U$-$B)$^2$ values. We divided our sample of galaxies into multiple $\MB$ and (U$-$B) subsamples and, for each subsample, compared the average SFR obtained from the \citet{Mostek12} calibration with that obtained from the stack of the rest-frame 1.4~GHz continuum luminosity density. We find that the difference in SFRs from the two approaches (as a function of colour and $\MB$) is consistent with the SFR scatter of 0.2~dex obtained by \citet{Mostek12}.}.

\subsection{The Stacking Analysis}
\label{sec:stacking}
We estimate the average \hi\ mass and the average SFR of subsamples of galaxies by stacking, respectively, the \hii\ line luminosities and the rest-frame 1.4~GHz continuum luminosities. The procedures used in stacking the \hii\ emission signals and the rest-frame 1.4~GHz continuum emission signals are described in detail in \citet{Chowdhury22a,Chowdhury22b}. We provide here, for completeness, a brief review of the procedures.

 The stacked \hii\ spectral cube of a given subsample of galaxies was computed by taking a weighted-average of the individual \hii\ subcubes, in luminosity-density units, of the galaxies in the subsample. { During the stacking analysis, each \hii\ subcube is treated as arising from a separate ``object''.} The weights were chosen to ensure that the redshift distributions of the different subsamples are identical; the specific choices of weights for the different stacks are discussed in Section~\ref{ssec:hi_stellar} and Section~\ref{ssec:hi_sSFR}. For each subsample, we then fitted a second-order polynomial to the spectra at each spatial pixel of the stacked \hii\ cube, and subtracted this out to obtain a residual cube; the polynomial fit was performed after excluding spectral channels in the velocity range $\pm250$~\kmps. For each subsample, the RMS noise at each spatial and velocity pixel of the stacked \hii\ cube was obtained via Monte Carlo simulations \citep{Chowdhury22b}. Finally, for each subsample, the average \hi\ mass was obtained from the measured velocity-integrated \hii\ line luminosity.\footnote{ Note that the quoted average \hi\ masses of the different subsamples in this Letter do not include the mass contribution of Helium.} The velocity integral was carried out over a contiguous range of central velocity channels containing emission at  $\geq1.5\sigma$ significance, after smoothing the stacked \hii\ subcubes to a velocity resolution of 90~\kmps. 

The average SFR of each subsample was computed by stacking the rest-frame 1.4~GHz luminosity density of the galaxies in the subsample \citep[e.g.][]{White07,Chowdhury22a}. We used the GMRT 655~MHz radio-continuum images of the DEEP2 subfields to extract subimages around each of the 11,419 galaxies of the full sample. We convolved all subimages to an uniform spatial resolution of 40~kpc, regridded them to a uniform grid with $5.2$~kpc pixels spanning $\pm260$~kpc, and converted the flux-density values (in Jy) to rest-frame 1.4~GHz luminosity density values (in W~Hz$^{-1}$), assuming a spectral index of $\alpha=-0.8$ \citep{Condon92}, with $S_{\nu} \propto \nu^{\alpha}$. The stacked rest-frame 1.4~GHz luminosity density of a subsample of galaxies was computed by taking a weighted-median of the individual subimages, with the weights being the same as those used during the \hii\ stacking of the subsample. Finally, the stacked rest-frame 1.4~GHz continuum luminosity density of a subsample of galaxies is converted to an estimate of the average SFR of the subsample, using the relation  SFR~$(\Msun/\textrm{yr}) = 3.7 \times 10^{-22} \times {\rm L_{1.4 GHz}\ (W~Hz^{-1})}$ \citep{Yun01}. The errors on our measurements of the average SFRs include both the statistical uncertainty and a 10$\%$ flux-scale uncertainty \citep{Chowdhury22b}.

\section{Results and Discussion}
\subsection{\hi\ Mass as a Function of Stellar Mass}
\label{ssec:hi_stellar}
\begin{figure} 
    \centering
    \includegraphics[width=0.95\linewidth]{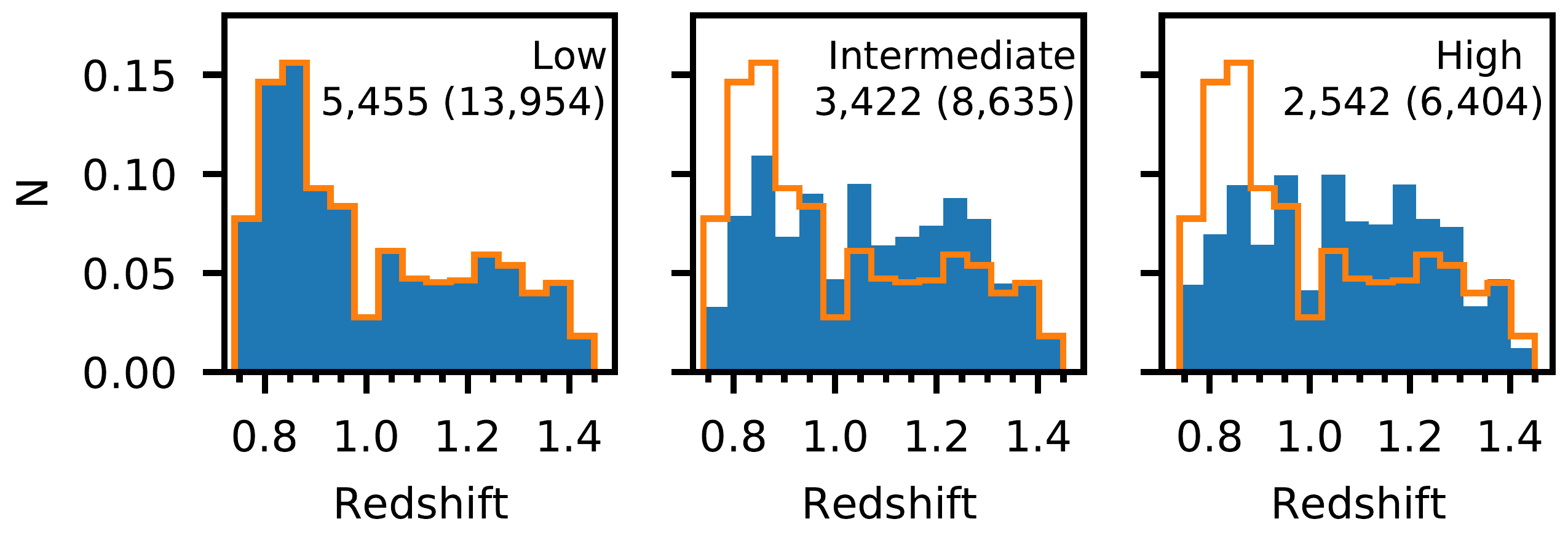}   
\caption{{ The redshift distributions of the three stellar-mass subsamples. The blue histograms show, for each stellar-mass subsample, the number of \hii\ subcubes in different redshift intervals (N), obtained after normalising by the total number of subcubes in the corresponding subsample.} The \hii\ subcubes of each stellar-mass subsample were assigned weights such that each effective redshift distribution is identical to the redshift distribution of the low stellar-mass subsample (orange lines). The number of galaxies in the subsample is indicated in each panel, with the number of \hii\ subcubes shown in parentheses.}
 \label{fig:redshiftdistMs}
\end{figure}

 We divide our sample of 11,419 galaxies (28,993 \hii\ subcubes) into three stellar-mass subsamples with $1.0 \times 10^{9}~\Msun<\Ms\le6.0 \times 10^{9}\ \Msun$ (``Low"), $6.0 \times 10^{9}\ \Msun<\Ms\le 1.3 \times 10^{10}\ \Msun$ (``Intermediate"), and $\Ms > 1.3 \times 10^{10}\ \Msun$ (``High")\footnote{ The stellar-mass ranges of the three  subsamples were chosen such that a clear ($\geq 4\sigma$) detection of the stacked \hii\ emission signal is obtained for each subsample. However, we emphasise that the conclusions of this Letter do not depend on the exact choice of the stellar-mass bins.} The number of galaxies and \hii\ subcubes in each subsample are provided in Table~\ref{tab:Mssubsamples}.  The redshift distributions of the three stellar-mass subsamples are different (see  Figure~\ref{fig:redshiftdistMs}). We correct for this difference by assigning weights to each \hii\ subcube such that the redshift distribution of each stellar-mass subsample is effectively identical. Specifically, the weights ensure that the effective redshift distributions of the intermediate- and high-stellar-mass subsamples are identical to that of the low-stellar-mass subsample; the mean redshift of the final redshift distribution is $\langle z \rangle=1.01$. We use these weights while computing all average quantities for the three stellar-mass subsamples. 

\begin{figure}

\centering
\includegraphics[width=\linewidth]{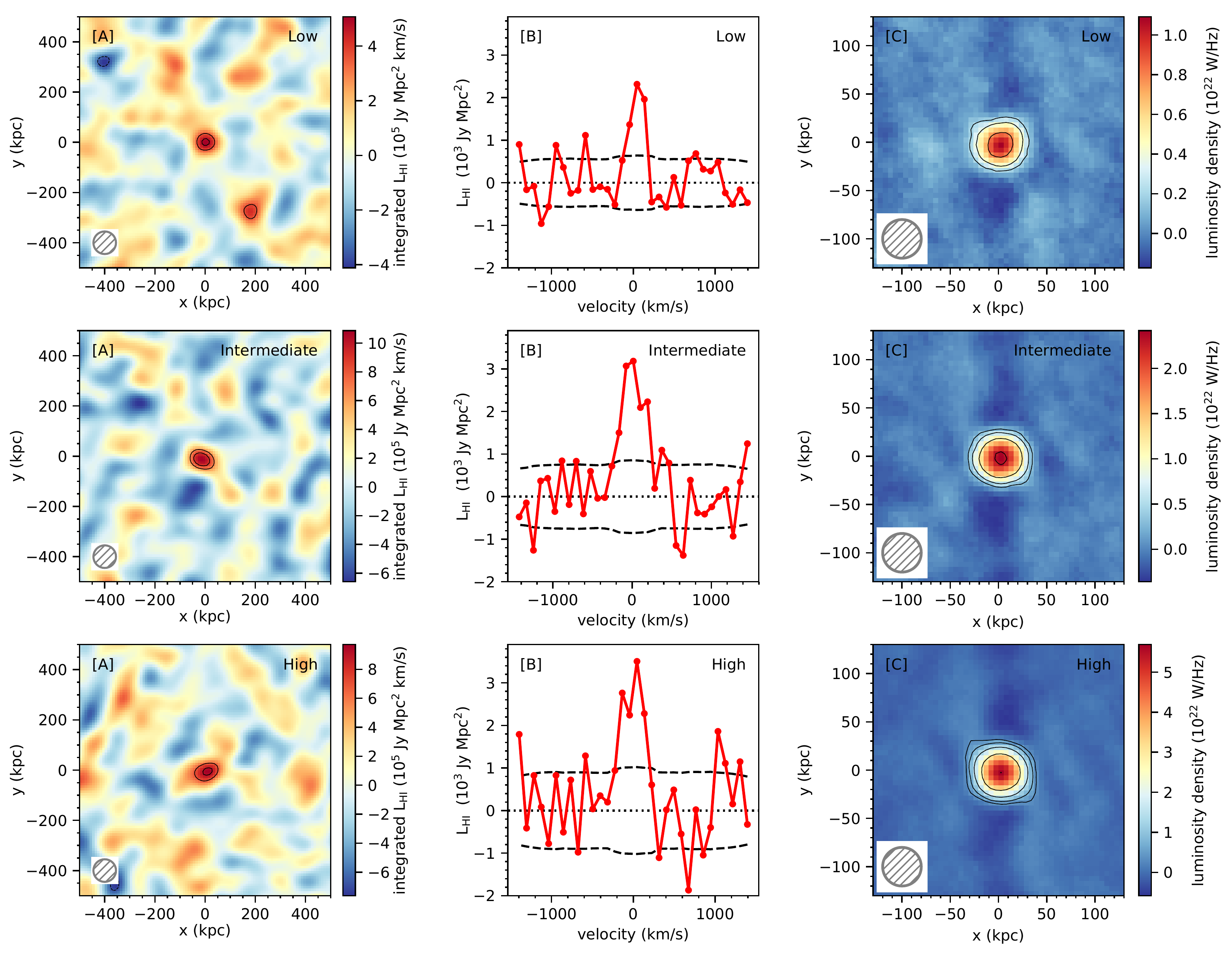}
\caption{The average \hii\ emission signal and the average rest-frame 1.4~GHz continuum emission from star-forming galaxies in the three stellar-mass subsamples. Panels~[A] show the average \hii\ emission images of galaxies of the three stellar-mass subsamples. The \hii\ subcubes of each subsample were assigned weights such that their effective redshift distributions are identical. The circle on the bottom left of each panel indicates the 90-kpc spatial resolution of the images. The contour levels are at $-3.0\sigma$ (dashed), $+3.0\sigma$, and $+4.0\sigma$ significance.  Panels~[B] show the average \hii\ emission spectra of the three stellar-mass subsamples. The $\pm1\sigma$ errors on the stacked \hii\ spectra are indicated by the dashed black curves. We clearly detect the stacked \hii\ emission signals in all three subsamples. Panels~[C] show the average rest-frame 1.4~GHz luminosity density of the galaxies in the three stellar-mass subsamples. The contour levels are at $5\sigma,\ 10 \sigma,\ 20\sigma,\ 40\sigma,\ {\rm and} \ 80\sigma$ statistical significance. The circle at the bottom left of each panel indicates the 40~kpc resolution of the images.}
\label{fig:msStacks}
\end{figure}

We separately stacked the \hii\ emission and the rest-frame 1.4~GHz continuum emission of the galaxies in the three stellar-mass subsamples, following the procedures of Sections~\ref{sec:stacking}. Figure \ref{fig:msStacks} shows the stacked \hii\ emission images, the stacked \hii\ spectra, and the stacked rest-frame 1.4~GHz continuum images of the three subsamples. We obtain clear detections of the average \hii\ emission signal in all three cases, at $4.2-4.9\sigma$ statistical significance. 
We also detect the stacked rest-frame 1.4~GHz continuum emission at high significance ($>28\sigma$) in all three subsamples. The average \hi\ masses and the average SFRs of galaxies in the three subsamples are listed in Table~\ref{tab:Mssubsamples}. We find that the average SFR and the average stellar mass of the galaxies in the three subsamples are in excellent agreement with the star-forming main sequence at $z\approx1$ \citep[see Table~\ref{tab:Mssubsamples};][]{Whitaker14,Chowdhury22a}.

\begin{table}
\centering
\begin{tabular}{|l|c|c|c|}
\hline
\hline
    \,  & Low & Intermediate & High \\
    \hline
    Stellar Mass Range ($\times 10^{9}\ \Msun$) & $1.0-6.0$  &  $6.0-13$  & $13-240$ \\
    \hline
     Number of \hii\ Subcubes & 13,954 & 8,635 & 6,404\\
     \hline
    Number of Galaxies & 5,455 & 3,422 & 2,542\\
     \hline
    Average Redshift   & 1.01 & 1.01 & 1.01 \\
       \hline
    Average Stellar Mass ($\times 10^{9}\ \Msun$) & $3.3$ & $8.9$   & $25.9$  \\
   
\hline
    Average H{\sc i} Mass  ($\times 10^{9}\ \Msun$) & $9.5\pm2.2$ & $20.3\pm4.1$ & $18.2\pm4.3$ \\
\hline
    Average SFR ($\Msun\textrm{yr}^{-1}$)& $4.04\pm0.43 $ & $8.95\pm0.92$ & $21.1\pm2.1$\\
\hline 
    Main-sequence SFR ($\Msun\textrm{yr}^{-1}$)  & 3.8 & 8.9 & 17.3 \\
\hline
    H{\sc i} depletion timescale (Gyr) & $2.35\pm0.61$ & $2.27\pm0.52$ & $0.86\pm0.22$ \\
\hline
     \hline
\end{tabular}

\caption{Average properties of galaxies in the three stellar-mass subsamples. For each subsample, the rows are (1)~the range of stellar masses, in units of $10^9 \ \Msun$, (2)~the number of \hii\ subcubes, (3)~the number of galaxies, (4)~the average redshift of the galaxies, (5) the average stellar mass of the galaxies, (6)~the average H{\sc i} mass of the galaxies, measured from the stacked \hii\ emission spectra of Figure~\ref{fig:msStacks}[B], (7)~the average SFR of the galaxies, measured using the stacked rest-frame 1.4~GHz luminosity densities of Figure~\ref{fig:msStacks}[C],  (8) the expected SFR at this average stellar mass, for the star-forming main sequence at $z\approx1$ \citep{Whitaker14}, and (9) the characteristic \hi\ depletion timescale, $\langle \MHI \rangle/\langle {\rm SFR} \rangle$.  Note that all quantities are weighted averages, with weights such that the redshift distributions of the three stellar-mass subsamples are identical.}
\label{tab:Mssubsamples}
\end{table}

\begin{table}
\centering
\begin{tabular}{|l|c|c|c|}
\hline
\hline
    \,  & Low & Intermediate & High \\
    \hline
    Stellar Mass Range ($\times 10^{9}\ \Msun$) & $1.0-6.0$  &  $6.0-13$  & $13-240$ \\ \hline
    Average Stellar Mass ($\times 10^{9}\ \Msun$) & $3.3$ & $8.9$   & $25.9$  \\
   
\hline
   {Average H{\sc i} Mass  ($\times 10^{9}\ \Msun$)}  & & & \\
      $z\approx0$ & $2.7\pm0.2$ & $4.5\pm0.4$ & $5.9\pm0.4$ \\
     $z\approx1$ & $9.5\pm2.2$ & $20.3\pm4.1$ & $18.2\pm4.3$ \\


\hline
    Average SFR ($\Msun\textrm{yr}^{-1}$)&  &  & \\
     $z\approx0$ & $0.44 \pm 0.03$ & $0.88\pm0.07$ & $1.83\pm0.15$ \\
     $z\approx1$  & $4.07\pm0.43 $ & $8.93\pm0.86$ & $21.1\pm2.1$\\
      
\hline
    H{\sc i} depletion timescale (Gyr) &  &  & \\

    $z\approx0$ & $6.11\pm0.48$  & $5.12\pm0.42$ &  $3.23\pm0.29$ \\
     $z\approx1$  & $2.33\pm0.60$ & $2.27\pm0.51$ & $0.86\pm0.26$ \\
\hline
     \hline
\end{tabular}

\caption{A comparison of the average \hi\ properties of blue star-forming galaxies at $z\approx1$ with those of blue galaxies in the local Universe. For galaxies in each of the three stellar-mass subsamples at both $z\approx0$ and $z\approx1$, the rows are (1)~the range of stellar masses, in units of $10^9 \ \Msun$, (2) the average stellar mass, (3)~the average H{\sc i} mass, (4)~the average SFR,  (5) the characteristic \hi\ depletion timescale, $\langle \MHI \rangle/\langle {\rm SFR} \rangle$. The \hi\ and stellar  properties of the $z\approx0$ subsamples are derived from the xGASS survey \citep{Catinella18}, using appropriate weights (see main text for details). The errors on the local Universe measurements were derived using bootstrap resampling with replacement.}
\label{tab:localUnivComp}
\end{table}

We use the extended GALEX Arecibo SDSS survey \citep[xGASS; ][]{Catinella18} to compare our measurements of the \hi\ properties of star-forming galaxies at $z\approx1$ to those of galaxies in the local Universe. The xGASS Survey used the Arecibo telescope to measure the \hi\ masses of a stellar-mass-selected sample of galaxies with $\Ms>10^{9} \ \Msun$ at $z = 0.01-0.05$. { The stellar masses and SFRs of the xGASS galaxies used in this work were obtained from the publicly available catalogue of the ``xGASS representative sample". The stellar masses in this catalogue are from \citet{Kauffmann03} and \citet{Brinchmann04}, while the SFRs were computed using a combination of Galex near-ultraviolet (NUV) and WISE mid-infrared (MIR) data or via spectral energy distribution fits for galaxies for which MIR data were not available \citep{Catinella18}.} 

The main sample of the GMRT-CAT$z1$ survey consists of blue, star-forming galaxies at $z=0.74-1.45$. In order to ensure a fair comparison between the \hi\ properties of the GMRT-CAT$z1$ galaxies and those of the xGASS galaxies, we restrict to blue galaxies,  with NUV$-$r$<4$, in the xGASS sample. We divide the xGASS galaxies into three stellar-mass subsamples, using the same ``Low'', ``Intermediate'', and ``High'' stellar-mass ranges as for the DEEP2 galaxies. Further, for each xGASS subsample, we use weights to ensure that the stellar-mass distribution within the subsample is effectively identical to that of the corresponding (Low, Intermediate, or High) subsample at $z \approx 1$. { In passing, we note that the average \hi\ mass of xGASS galaxies in the three stellar-mass subsamples obtained using a cut in the SFR-$\Ms$ plane to select main-sequence galaxies is consistent with the values obtained by selecting blue galaxies with NUV$-$r$<4$.}

The average \hi\ masses of the blue xGASS galaxies in the three stellar-mass sub-samples are listed in Table~\ref{tab:localUnivComp}; the errors on the averages were computed using bootstrap resampling with replacement. The table also lists, for comparison, the GMRT-CAT$z$1 measurements of the average \hi\ masses of blue galaxies in the same stellar-mass subsamples at $z\approx1$. We find that, across the stellar-mass range $10^9 - 2.4 \times 10^{11}~\Msun$, the average \hi\ mass of the $z \approx 1$ galaxies is higher than that of local Universe galaxies, by a factor of $\approx3.1-4.5$.




We determined the $\MHI-\Ms$ relation at $z\approx1$ by fitting a power-law relation to our measurements of the average \hi\ mass of blue star-forming galaxies in the three stellar-mass subsamples at $z\approx1$, following the procedures in Appendix~\ref{sec:fitting}. We find that the $\MHI-\Ms$ relation for main-sequence galaxies at $z\approx1$ is

\begin{equation}
        \log \left[\MHI/\Msun\right]= (0.32\pm0.13)  \log\left[{\textrm{M}_{*,10}}\right]+(10.183\pm0.056)  \;,
        \label{eqn:MHI_Ms}
\end{equation}
where ${\textrm{M}}_{*,10}=\Ms/10^{10}~\Msun$. In order to compare the $\MHI-\Ms$ relation of blue star-forming  galaxies at $z\approx1$ to that of blue star-forming galaxies at $z\approx0$, we also fitted a power-law relation, using the procedures of Appendix~\ref{sec:fitting}, to the measurements of $\langle \MHI \rangle$ in blue xGASS galaxies in the three stellar-mass subsamples of Table~\ref{tab:localUnivComp}, with stellar-mass distributions identical to those of the subsamples of galaxies at $z\approx1$. We find that the best-fit $\MHI-\Ms$ relation for blue galaxies at $z\approx0$ is $\log \left[\MHI/\Msun\right]= (0.38\pm0.05)  \log\left[{\Ms}_{,10}\right]+(9.634\pm0.019)$.\footnote{We note that the $\MHI-\Ms$ relation for blue xGASS galaxies obtained by fitting to the $\langle \MHI \rangle$ values in the three stellar-mass subsamples is consistent with that obtained by fitting to $\langle \MHI \rangle$ values in small $\Ms$ bins, separated by 0.1~dex.}

\begin{figure}
    \centering
         \includegraphics[width=\linewidth]{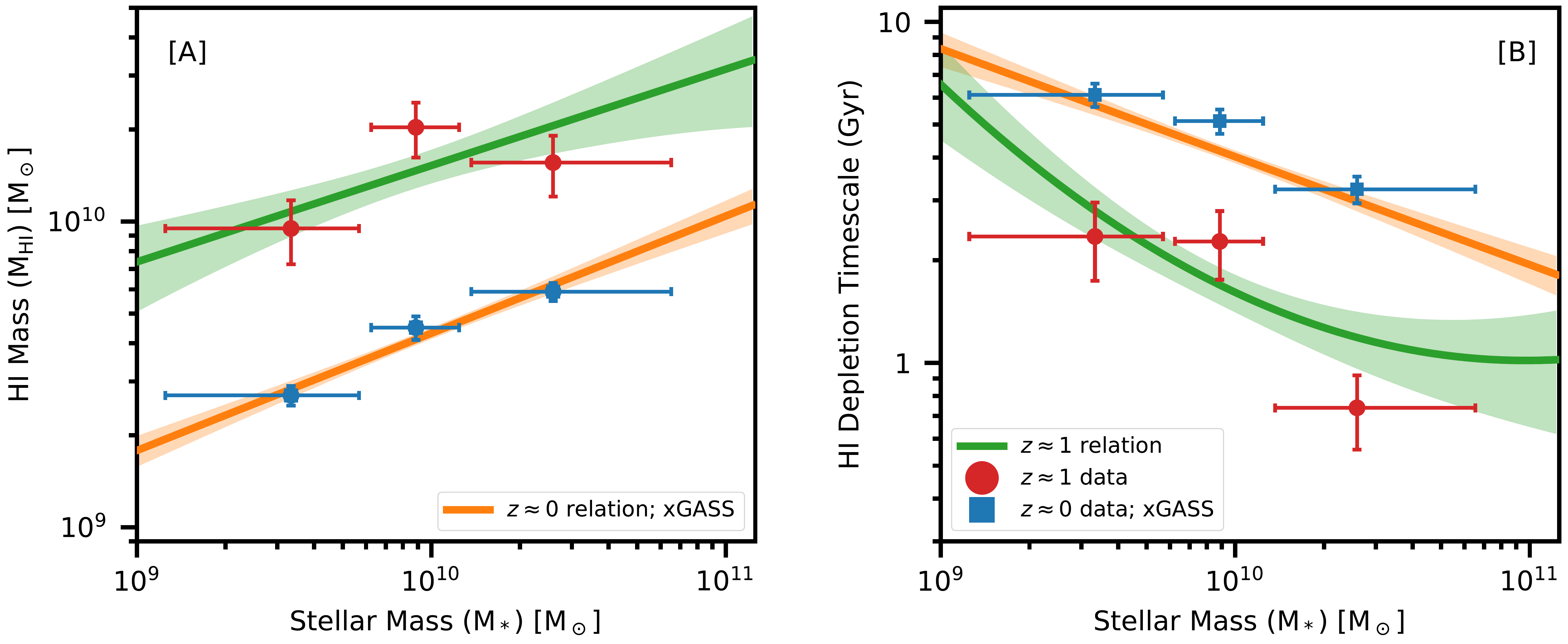}
     \caption{The \hi\ properties of star-forming galaxies at $z\approx1$, as a function of their stellar masses. The red circles in panels [A] and [B] show, respectively, our measurements of the average HI mass, $\langle\MHI\rangle$, and the characteristic \hi\ depletion timescale, $\langle\tdephi\rangle=\langle\MHI\rangle/\langle\textrm{SFR}\rangle$, for star-forming galaxies at $z\approx1$ in the three stellar-mass subsamples of Figure~\ref{fig:msStacks}. The blue squares indicate the same quantities for the blue xGASS galaxies in three $\Ms$ subsamples with stellar-mass distributions identical to those of the three subsamples at $z\approx1$.  The $\MHI-\Ms$ relation at $z\approx1$, derived by fitting a power-law relation to our measurements of the average \hi\ mass in the three stellar-mass subsamples, is shown as the green line in Panel~[A], with the green shaded region showing the $1\sigma$ error on the relation. Panel~[A] also shows the  $\MHI-\Ms$ relation for blue galaxies at $z\approx0$ (orange line), obtained by fitting a power-law relation to the average \hi\ mass of xGASS galaxies in the three $\Ms$ subsamples. In Panel~[B], the green curve shows the $\tdephi-\Ms$ relation at $z\approx1$, derived by combining our estimate of the $\MHI-\Ms$ relation at $z\approx1$ with the equation describing the star-forming main sequence at $z\approx1$ \citep{Whitaker14}; the green shaded region shows the $1\sigma$ error on the relation. { The orange line in panel~[B] shows an estimate of the $\tdephi-\Ms$ relation at $z \approx 0$ derived in a similar manner, by combining the $\MHI-\Ms$ relation at $z\approx0$ with the equation describing the star-forming main sequence at $z\approx0$ \citep{Whitaker12}.}  The figure shows that blue star-forming galaxies at $z\approx1$, with stellar masses in the range $\Ms\approx10^9-10^{11}~\Msun$, have  $\approx3-4$ times more \hi\ than blue galaxies at $z\approx0$, but have far lower characteristic depletion timescales, by a factor of $\approx2-4$.  }
    \label{fig:HI_Ms}
\end{figure}

Figure~\ref{fig:HI_Ms}[A] shows the  $\MHI-\Ms$ relations for blue star-forming galaxies at $z\approx1$ and $z\approx0$. We find no statistically significant evidence for an evolution in the slope of the $\MHI-\Ms$ relation from $z \approx 1$ to $z \approx 0$. However, we find clear evidence that the relation has shifted downwards from $z\approx1$ to $z\approx0$. Specifically, our measurements show that the $\MHI-\Ms$ relation of blue star-forming galaxies at $z\approx1$ lies a factor of $3.54\pm0.48$ above the local Universe relation.

{ In passing, we emphasize that the $\MHI-\Ms$ relations of this Letter, at both $z\approx0$ and $z\approx1$, were obtained by fitting a relation to measurements of $\langle\MHI\rangle$ in three stellar-mass subsamples. This approach is different from that typically followed for galaxies at $z\approx0$, where the $\MHI-\Ms$ relation is obtained by fitting to estimates of $\langle\log\MHI\rangle$ in multiple stellar-mass subsamples \citep[e.g.][]{Saintonge22}. The difference arises from the fact that the averaging in a stacking analysis is carried out on the \hi\ masses themselves, rather than on the logarithm of the \hi\ masses; in general, the logarithm of the average value of a given quantity is not the same as the average of the individual logarithms \citep[e.g.][]{Brown15}. Care must hence be taken when comparing scaling relations obtained from simulations with those obtained from stacking analyses such as the present work, or when comparing the scaling relations from stacking analyses with those based on direct measurements of $\MHI$, and hence on estimates of $\langle\log\MHI\rangle$. Specifically, the scaling relations obtained from the stacking analysis yield the mean \hi\ mass at a given stellar mass. Conversely, for a log-normal distribution of \hi\ masses, the scaling relations obtained from direct measurements yield the median HI mass at a given stellar mass. Further, again for a log-normal distribution of \hi\ masses with scatter $\sigma$, $\langle\log\MHI\rangle$ = $\log\langle\MHI\rangle-\frac{\ln10}{2}\sigma^2$. Assuming that the scatter of the scaling relation at $z\approx1$ is independent of $\Ms$ and that it is equal to the scatter of $0.4$~dex measured at $z = 0$ \citep{Catinella18}, the ``direct" $\MHI-\Ms$ relation would be offset downward from Equation~\ref{eqn:MHI_Ms} by 0.184~dex. }

\subsection{The \hi\ Depletion Timescale as a Function of Stellar Mass}

The availability of cold gas regulates the star-formation activity in a galaxy. The \hi\ depletion timescale ($\tdephi$), defined as the ratio of the \hi\ mass of the galaxy to its SFR, quantifies the approximate timescale for which the galaxy can sustain its current SFR, in the absence of accretion of fresh \hi\ from the CGM. In other words, accretion of gas from the CGM on a timescale of $\approx \tdephi$ is required to sustain the current star-formation activity of the galaxy.

We define the ``characteristic'' \hi\ depletion timescale of a sample of galaxies as $\langle\tdephi\rangle\equiv\langle\MHI\rangle/\langle{\rm SFR}\rangle$.  We combined the average SFRs of galaxies in the three subsamples with their average \hi\ masses to estimate the characteristic \hi\ depletion timescale, $\langle\tdephi\rangle\equiv\langle\MHI\rangle/\langle{\rm SFR}\rangle$, of galaxies at $z\approx1$, as a function of their average stellar masses. Table~\ref{tab:Mssubsamples} lists the $\langle\tdephi\rangle$ values  of the galaxies in the three stellar-mass subsamples at $z \approx 1$, while the estimates of 
  $\langle\tdephi\rangle$ are plotted against the average stellar mass in Figure~\ref{fig:HI_Ms}[B]. For comparison, the figure also shows the characteristic \hi\ depletion timescale of the xGASS galaxies in the same three stellar-mass subsamples, while Table~\ref{tab:localUnivComp} compares the values of $\langle\tdephi\rangle$ for the galaxies at $z\approx0$ and $z \approx 1$. We find that the characteristic \hi\ depletion timescale of blue star-forming galaxies at $z\approx1$ is $\approx2-4$ times lower than that of similar galaxies with the same stellar mass distribution at $z\approx0$. 

In passing, we note that the ``characteristic'' \hi\ depletion timescale, $\langle\MHI\rangle/\langle{\rm SFR}\rangle$, for a sample of galaxies may be different from the average of the depletion timescales of the individual galaxies, $\langle \MHI / SFR \rangle$. Indeed, for the xGASS galaxies, we find that the $\langle \MHI/{\rm SFR} \rangle$ values in the three stellar-mass subsamples are higher than the corresponding $\langle \MHI \rangle / \langle {\rm SFR} \rangle$ values by factors of $\approx 1.2-1.6$. However, this does not affect the results of this \emph{Letter} because we consistently compare the characteristic depletion timescales of the different galaxy subsamples, at both $z\approx 1$ and $z \approx 0$.

We obtained the $\tdephi-\Ms$ relation at $z\approx1$ by combining our estimate of the $\MHI-\Ms$ relation at $z\approx1$ (Equation~\ref{eqn:MHI_Ms}) with a relation for the star-forming main sequence at $z\approx1$ from \citet{Whitaker14}. These authors provide best-fitting relations to the star-forming main sequence for the redshift ranges $z=0.5-1.0$ and $z=1.0-1.5$; we interpolated the best-fit parameters between the two redshift intervals to find that the main-sequence relation at $z\approx1$ is $\log \left[{\rm SFR}/(\Msun {\rm yr}^{-1})\right]= 0.976 + 0.720 \log\left[{\Ms}_{,10}\right] - 0.205 \log\left[{\Ms}_{,10}\right]^2 $. Combining this relation with the $\MHI-\Ms$ relation of Equation~\ref{eqn:MHI_Ms}, we find that the $\tdephi-\Ms$ relation for main-sequence galaxies at $z\approx1$ is\footnote{We note that the uncertainties on the $\tdephi-\Ms$ relation of Equation~\ref{eqn:tdepHI} are dominated by the uncertainties on the $\MHI-\Ms$ relation at $z\approx1$, with relatively little contribution from the uncertainties in the main-sequence relation of \citet{Whitaker14}. The errors on the parameters in Equation~\ref{eqn:tdepHI} were hence obtained by ignoring the uncertainties in the main-sequence relation.}: 

\begin{equation}
\log \left[\tdephi/{\rm Gyr}\right]= (0.207\pm0.056) + (-0.40\pm 0.13) \log\left[{\Ms}_{,10}\right]+ 0.205 \log\left[{\Ms}_{,10}\right]^2
\label{eqn:tdepHI}
\end{equation}

We emphasise that Equation~\ref{eqn:tdepHI}  was not obtained by fitting a relation to our measurements of the characteristic \hi\ depletion timescale in the three $\Ms$ subsamples. However, Figure~\ref{fig:HI_Ms}[B] shows that our measurements of $\langle\tdephi\rangle$ in the three subsamples are consistent with the $\tdephi-\Ms$ relation of Equation~\ref{eqn:tdepHI}.

   Overall, we find that blue star-forming galaxies at $z\approx1$, with stellar masses in the range $\Ms\approx10^9- 2.4 \times 10^{11}~\Msun$\ have larger \hi\ reservoirs than those of blue galaxies at $z\approx0$, by a factor of $3.54\pm0.48$. However, the evolution of the star-forming main-sequence by a factor of $\approx10$ from $z\approx0$ to $z\approx1$ \citep[e.g.][]{Whitaker14} implies that the characteristic \hi\ depletion timescales of blue star-forming galaxies at $z\approx1$ are lower, by factors of $\approx 2-4$, than those of local galaxies. The results of this \emph{Letter} thus extend the findings of the earlier GMRT \hii\ stacking studies  \citep{Chowdhury20,Chowdhury21,Chowdhury22a,Chowdhury22b} that blue star-forming galaxies at $z\approx1$ have a large average \hi\ mass but a short characteristic \hi\ depletion timescale to the entire stellar mass range $\Ms\approx10^9-2.4 \times 10^{11}~\Msun$.

\subsection{The \hi\ Fraction as a Function of the Specific SFR}
\label{ssec:hi_sSFR}

The \hi\ fractions ($\fhi \equiv \MHI/\Ms$) of galaxies in the local Universe and their specific SFRs ($\textrm{sSFR}\equiv\textrm{SFR}/\Ms$) are known to be correlated, with a scatter of $\approx0.5$~dex \citep{Catinella18}; this is one of the tightest atomic gas scaling relations at $z \approx 0$ \citep{Catinella18}. The locations of galaxies in the $\fhi-$sSFR plane are indicative of the efficiency with which their \hi\ is being converted to stars.  In this section, we investigate the redshift evolution of the relation between $\fhi$ and sSFR, for blue star-forming galaxies, from  $z \approx 1$ to $z \approx 0$.

\begin{figure}
    \centering
    \includegraphics[width=0.8\linewidth]{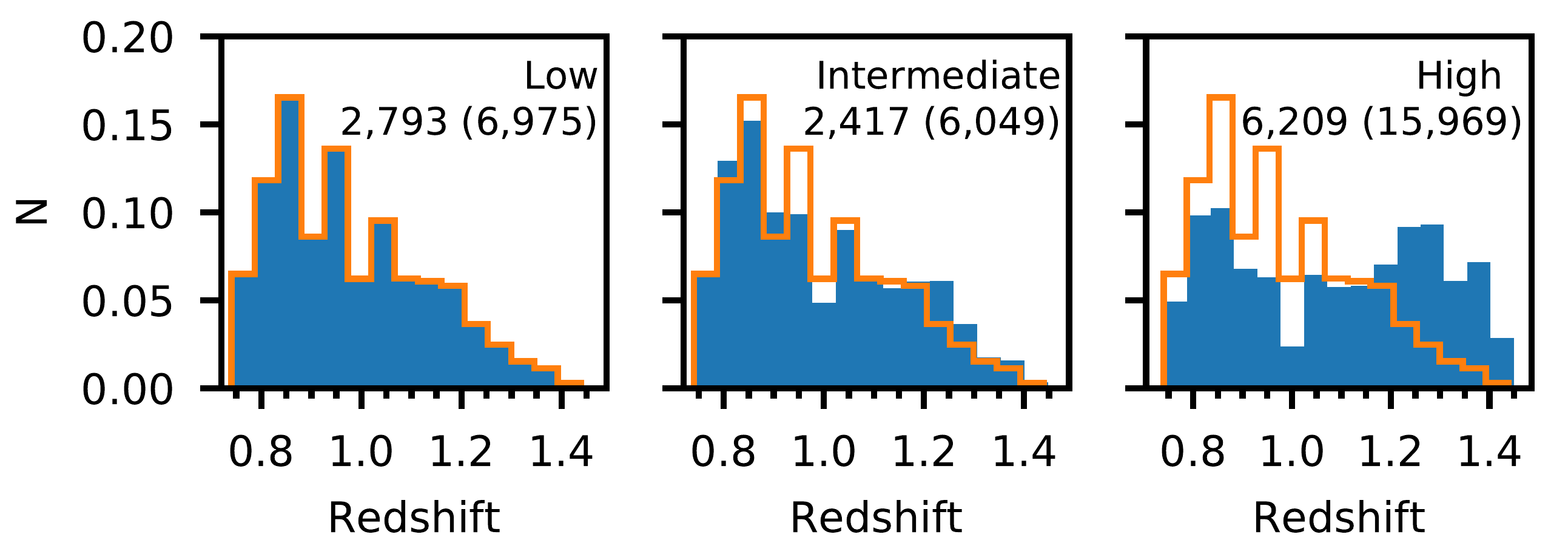}
\caption{{ The redshift distributions of the three sSFR subsamples. The blue histograms show, for each sSFR subsample, the number (N) of \hii\ subcubes normalized by the total number of subcubes in the corresponding subsample, for the different redshift intervals.} The \hii\ subcubes of each sSFR subsample were assigned weights such that each effective redshift distribution is identical to the redshift distribution of the low-sSFR subsample (orange lines). The total number of galaxies in the subsample is indicated in each panel, with the number of \hii\ subcubes shown in parentheses.}
\label{fig:redshiftdistsSFR}
\end{figure}

We divide our sample of 11,419 galaxies into three sSFR subsamples  with sSFR~$\le0.8~\textrm{Gyr}^{-1}$ (``Low''), $0.8 ~\textrm{Gyr}^{-1}<~$sSFR$~\le1.3~\textrm{Gyr}^{-1}$ (``Intermediate''), and  sSFR$~>1.3~\textrm{Gyr}^{-1}$ (``High'')\footnote{ The sSFR ranges of the three  subsamples were chosen such that a clear ($\geq 4\sigma$) detection of the stacked \hii\ emission signal is obtained for each subsample. However, we emphasise that the conclusions of this section do not depend on the exact choice of the sSFR bins.}. The numbers of galaxies and \hii\ subcubes in each subsample are listed in Table~\ref{tab:sSFRsubsamples}, while the redshift distributions of the three sSFR subsamples are shown in  Figure~\ref{fig:redshiftdistsSFR}. The high-sSFR subsample contains a significantly larger number of galaxies at higher redshifts than the other two subsamples; this is primarily due to the redshift evolution of the star-forming main sequence within our redshift coverage, $z=0.74-1.45$ \citep[e.g.][]{Whitaker14}. We corrected for this difference in the redshift distributions of the subsamples by using weights such that the effective redshift distributions of the intermediate- and high-sSFR subsamples are identical to that of the low-sSFR subsample. We separately stacked the \hii\ subcubes of the galaxies in the three subsamples, following the procedures of Section~\ref{sec:stacking}, using the above weights to ensure that the redshift distributions of the three subsamples are identical.

Figure~\ref{tab:sSFRsubsamples} shows the stacked \hii\ emission images and the stacked \hii\ spectra of galaxies in the three sSFR subsamples. We obtain clear detections, with $\approx4.3-4.4\sigma$ statistical significance, of the average \hii\ emission signals from galaxies in the three subsamples. The average \hi\ mass and the ``characteristic'' \hi\ fraction, $\langle\fhi\rangle\equiv\langle\MHI\rangle/\langle\Ms\rangle$, of the galaxies in each subsample are listed in Table~\ref{tab:sSFRsubsamples}.

\begin{table}
\centering
\begin{tabular}{|l|c|c|c|}
\hline
\hline
    \, & Low & Intermediate &  High \\
    \hline
    sSFR Range ($\textrm{Gyr}^{-1}$) & $0.1-0.8$  &  $0.8-1.3$&  $1.3-4.2$  \\
    \hline
     Number of \hii\ Subcubes & 6,975 & 6,049 & 15,969\\
     \hline
    Number of Galaxies & 2,793 & 2,417 & 6,209 \\
     \hline
    Average Redshift   & 0.97 & 0.97  & 0.97 \\
     \hline
    Average sSFR ($\textrm{Gyr}^{-1}$) & $0.5$ & $1.1$ & $1.9$   \\
    \hline
    Average Stellar Mass ($\times 10^{9}\ \Msun$) & $20.6$ & $9.4$ & $4.5$  \\
    \hline
    Average \hi\ Mass  ($\times 10^{9}\ \Msun$) & $15.1\pm3.4$ & $16.4\pm3.7$ & $9.1\pm2.1$ \\
    \hline
    Characteristic \hi\ Fraction & $0.73\pm0.17 $ & $1.75\pm0.39$ & $2.02\pm0.47$ \\
    \hline\hline
\end{tabular}
\caption{Average properties of galaxies in the three sSFR subsamples. For each sSFR subsample, the rows are (1)~the range of sSFR values, in units of $\textrm{Gyr}^{-1}$, (2)~the number of \hii\ subcubes, (3)~the number of galaxies, (4)~the average redshift, (5) the average sSFR, (6) the average stellar mass, (7)~the average \hi\ mass, measured from the stacked \hii\ emission spectra of Figure~\ref{fig:sSFRStacks}, and (8)~the characteristic \hi\ fraction, $\fhi \equiv \langle \MHI \rangle / \langle \Ms \rangle$.   Note that all quantities are weighted averages, with weights such that the redshift distributions of the three sSFR subsamples are identical.}
\label{tab:sSFRsubsamples}
\end{table}

\begin{figure}

\centering
\includegraphics[width=\linewidth]{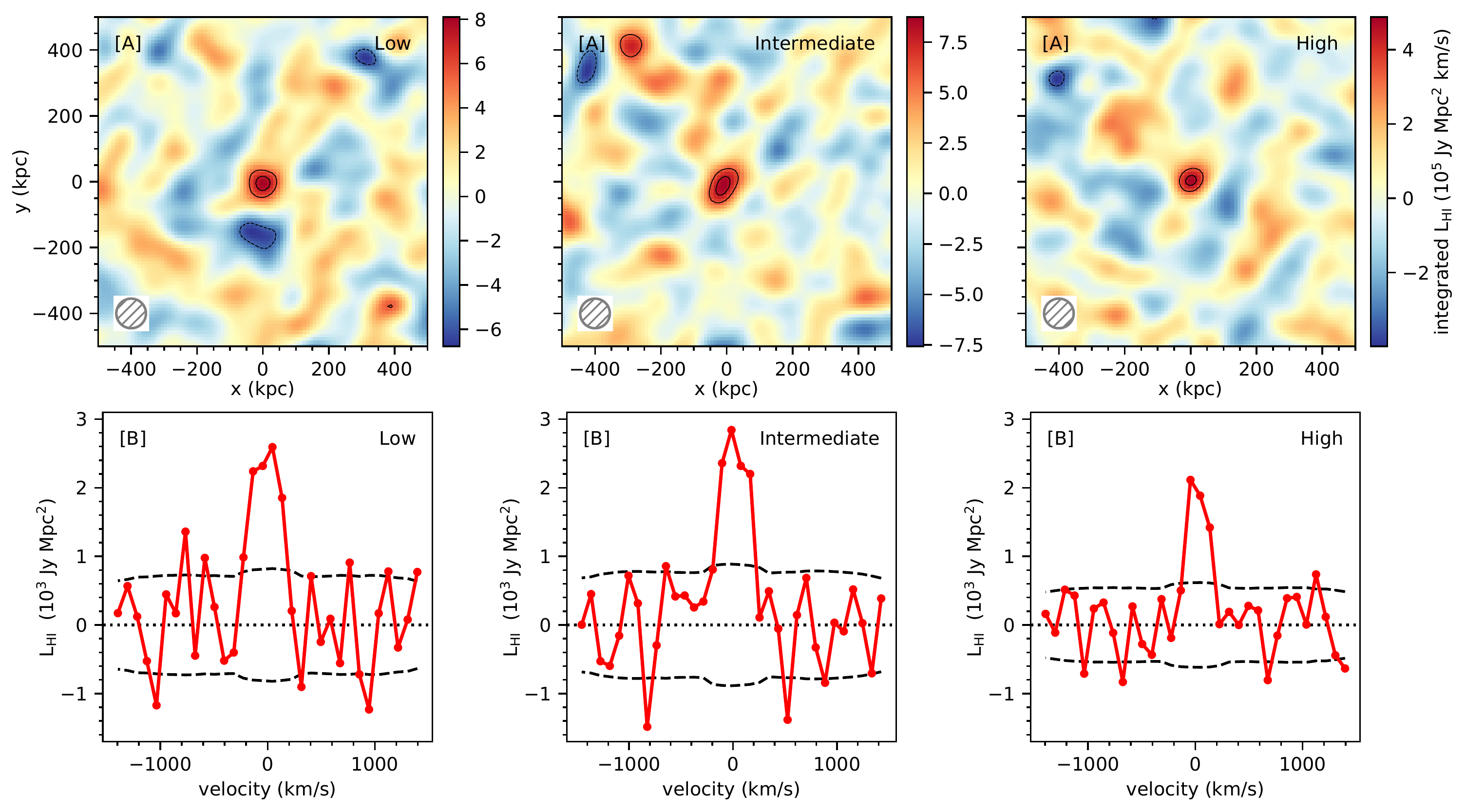}
\caption{
The average \hii\ emission signals from star-forming galaxies in the three sSFR subsamples. Panels~[A] show the average \hii\ emission images of the three sSFR mass subsamples. The circle on the bottom left of each panel indicates the 90-kpc spatial resolution of the images. The contour levels are at $-3.0\sigma$ (dashed), $+3.0\sigma$, and $+4.0\sigma$ significance.  Panels~[B] show the average \hii\ emission spectra of the same galaxies in the three sSFR subsamples. The $\pm1\sigma$ errors on the stacked \hii\ spectra are indicated with dashed black curves. We clearly detect the stacked \hii\ emission signals in all three subsamples. The \hii\ subcubes of each subsample were assigned weights such that their effective redshift distributions are identical.}
\label{fig:sSFRStacks}
\end{figure}

\begin{figure}
    \centering
     \includegraphics[width=0.6\linewidth]{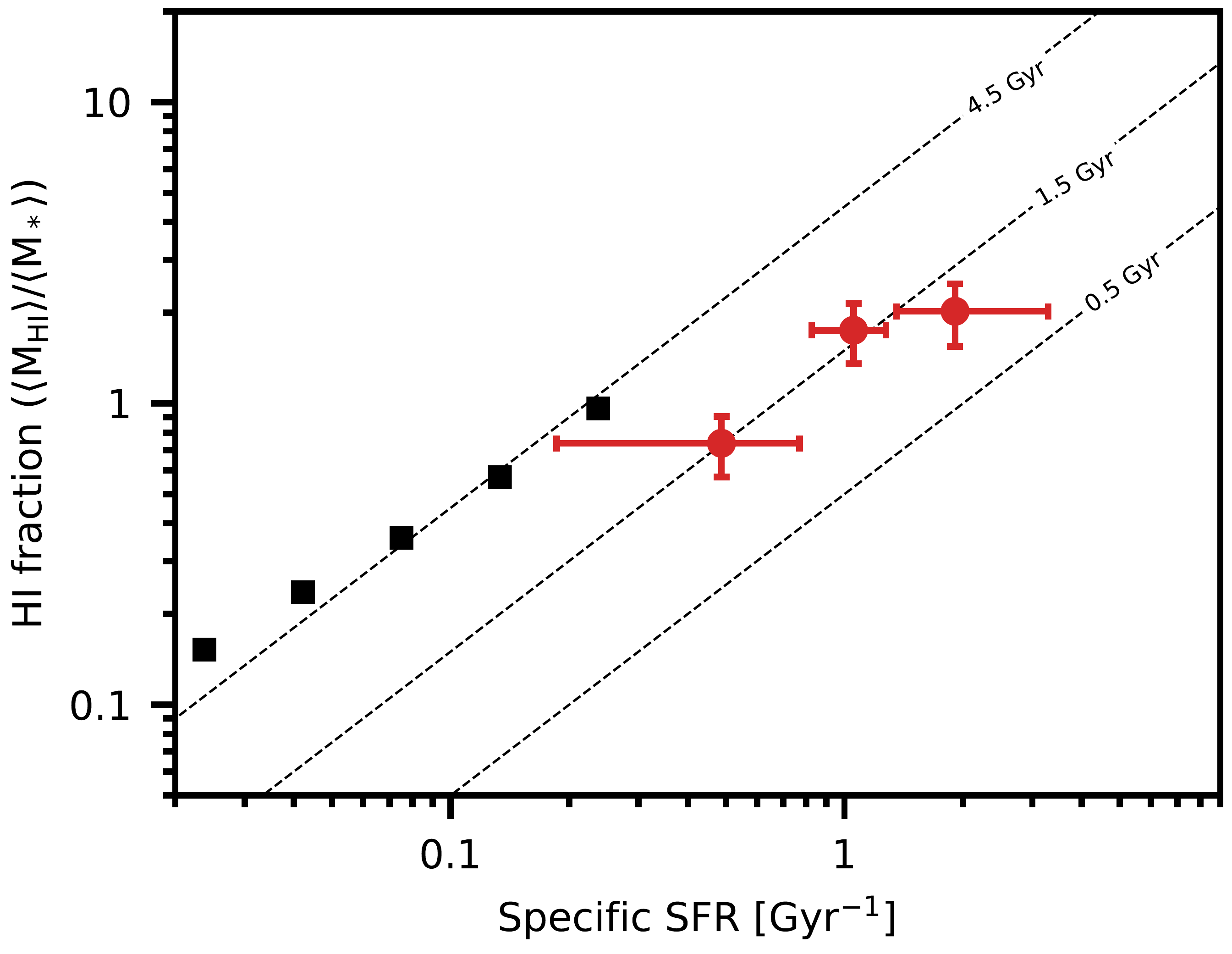}
    \caption{The characteristic \hi\ fractions of star-forming galaxies at $z\approx1$, as a function of their specific star-formation rates. The red circles show our measurements of the characteristic \hi\ fraction, $\langle \MHI \rangle$/$\langle \Ms \rangle$, of star-forming galaxies at $z\approx1$ in the three sSFR subsamples of Figure~\ref{fig:sSFRStacks}. The black squares  indicate the characteristic \hi\ fractions of blue xGASS galaxies at $z\approx0$ in multiple sSFR bins. The dashed lines show the loci of constant gas depletion timescales. The relation between the \hi\ fraction and the sSFR shows clear evolution from $z\approx1$ to $z\approx0$, with blue galaxies at $z\approx0$ having a characteristic \hi\ depletion timescale of $\approx4.5$~Gyr \citep[see also][]{Saintonge17} but those at $z\approx1$ having an \hi\ depletion timescale of just $\approx 1.5$~Gyr. }
    \label{fig:HIfrac_sSFR}
\end{figure}

  Our measurements of the characteristic \hi\ fraction of star-forming galaxies in the three sSFR subsamples at $z\approx1$ are shown in  Figure~\ref{fig:HIfrac_sSFR}; also shown for comparison are the characteristic \hi\ fractions of blue xGASS galaxies at $z\approx0$ \citep{Catinella18}. Note that the average sSFR of the DEEP2 galaxies in the low-sSFR subsample is $0.5~\textrm{Gyr}^{-1}$, while there are only 3 galaxies in the xGASS survey with sSFR $>0.5~\textrm{Gyr}^{-1}$. This is because the main sequence evolves between $z\approx1$ and $z\approx0$, with the sSFR of galaxies at a fixed stellar mass being $\approx10$ times higher at $z\approx1$ than at $z\approx0$ \citep[e.g.][]{Whitaker14}. 

 The straight lines in Figure~\ref{fig:HIfrac_sSFR} are the loci  of constant depletion timescales on the $\fhi-\Ms$ plane.  The characteristic \hi\ depletion timescale of main-sequence galaxies in the local Universe is $\approx4.5$~Gyr, with a large scatter around the mean \citep{Saintonge17}. Figure~\ref{fig:HIfrac_sSFR}  shows that the characteristic \hi\ fractions and the average sSFRs of blue xGASS galaxies at $z\approx0$ are consistent with the $\langle\tdephi\rangle=4.5$~Gyr line. However, it is clear from Figure~\ref{fig:HIfrac_sSFR} that star-forming galaxies at $z\approx1$ do not follow the $\fhi-$sSFR relation of local Universe galaxies. This is consistent with our earlier results \citep[e.g.][]{Chowdhury20,Chowdhury21} that blue star-forming galaxies at $z\approx1$ have a low characteristic \hi\ depletion timescale of $\approx1-2$~Gyr. This evolution of the $\fhi-$sSFR relation from $z\approx1$ to $z\approx0$ is different from the behaviour of the molecular component: the molecular gas depletion timescales in main-sequence galaxies are typically $\approx 0.5-0.7$~Gyr at $z \approx 0-1.5$, with no significant evidence for redshift evolution over $z \approx 0-1.5$ \citep[e.g.][]{Saintonge17,Genzel15}. 


 The short \hi\ depletion timescale of galaxies at $z\approx1$ (or, equivalently,  the high \hi\ star-forming efficiency) is indicative of a very efficient conversion of \hi\ to \htwo, which then directly fuels the high star-formation activity. The difference between local Universe galaxies (with massive \hi\ reservoirs but low star-forming efficiency) and star-forming galaxies at $z\approx1$ may lie in the typical \hi\ surface densities in the galaxies; a high \hi\ surface density is likely to be a requirement for efficient conversion of \hi\ to \htwo\  \citep[e.g.][]{Leroy08}. In other words, it appears that { the efficiency of conversion of \hi\ to stars is different at $z\approx1$, towards the end of the epoch of peak star-formation activity in the Universe, from that  at $z\approx0$, with the \hi\ in galaxies at $z\approx1$ being able to fuel star-formation far more efficiently than at $z\approx0$.} Measurements of the average \hi\ surface density profiles of the GMRT-CAT$z1$ galaxies would allow one to test this hypothesis.


\section{Summary}
In this \emph{Letter}, we report the first determinations of \hi\ scaling relations  of galaxies at $z\approx1$, measuring the \hi\ properties of blue star-forming galaxies at $z = 0.74-1.45$ as a function of stellar mass and sSFR, based on data from the GMRT-CAT$z$1 survey. We divided our main sample of 11,419 blue star-forming galaxies at $z\approx1$ into three stellar-mass subsamples and detected the stacked \hii\ emission signals from all three subsamples at $4.3-4.9\sigma$ significance. We fitted a power-law relation for the dependence of the average \hi\ mass on the average stellar mass, to obtain $\log \left[\MHI/\Msun\right]= (0.32\pm0.13)  \log\left[{\Ms}_{,10}\right]+(10.183\pm0.056)$. We compared the $\MHI-\Ms$ relation at $z\approx1$ to that for blue galaxies at $z\approx0$ to find that the slope of the $\MHI-\Ms$ relation at $z\approx1$ is consistent with that at $z\approx0$. However, we find that the $\MHI-\Ms$ relation at $z\approx1$ has shifted upwards from the relation at $z\approx0$, by a factor of $3.54\pm0.48$. We combined our measurements of the average \hi\ mass in the three stellar-mass subsamples with measurements of their average SFRs, obtained by stacking the rest-frame 1.4~GHz continuum emission, to obtain the characteristic \hi\ depletion timescale, $\langle\MHI\rangle/\langle\textrm{SFR}\rangle$, of the three subsamples.  We find that the characteristic \hi\ depletion timescale of blue star-forming galaxies at $z\approx1$, over the stellar mass range $\Ms\approx10^9-2.4 \times 10^{11}~\Msun$, is $\approx2-4$ times lower than that at $z\approx0$, for blue galaxies with similar stellar masses. We also divided the galaxies into three sSFR subsamples, obtaining detections of the stacked \hii\ emission signals in all three subsamples, at $\approx 4.3-4.4\sigma$ significance. We find that the $\fhi-$sSFR relation shows evidence for redshift evolution, with galaxies at $z\approx1$ having a lower characteristic \hi\ fraction, by a factor of $\approx3$, than what is expected from the extrapolation of the relation at $z\approx0$ to higher sSFR values. We thus find that star-forming galaxies at $z\approx1$ are able to convert their \hi\ reservoirs into stars with much higher efficiency than galaxies at $z\approx0$. This is unlike the situation for molecular gas, where the efficiency of conversion of molecular gas to stars in main-sequence galaxies shows no significant evolution  over $z\approx0-1.5$.

	\begin{acknowledgments}
	We thank the staff of the GMRT who have made these observations possible. The GMRT is run by the National Centre for Radio Astrophysics of the Tata Institute of Fundamental Research. We thank an anonymous referee for suggestions that improved this manuscript. NK acknowledges support from the Department of Science and Technology via a Swarnajayanti Fellowship (DST/SJF/PSA-01/2012-13). AC, NK, $\&$ JNC also acknowledge the Department of Atomic Energy for funding support, under project 12-R\&D-TFR-5.02-0700. 
	\end{acknowledgments}
    \software{      astropy \citep{astropy:2013}}
      
    \bibliography{bibliography.bib}

\bibliographystyle{aasjournal}	

\appendix

\section{Fitting Power-law Relations to Stacked Measurements}
\label{sec:fitting}
 We fitted a power-law relation of the form in Equation~\ref{eqn:MHI} to our measurements of the average \hi\ mass in the three stellar-mass subsamples to determine the dependence of the \hi\ mass of star-forming galaxies at $z \approx 1$ on their stellar mass. 

\begin{equation}
\log \left[\MHI (\alpha,\beta)/\Msun\right]= \alpha + \beta \log\left[{\Ms}_{,10}\right]  \; .
\label{eqn:MHI}
\end{equation}

 The fitting was done via a $\chi^2$ minimization, taking into account the stellar-mass distribution of the galaxies in each of the three subsamples. Specifically, for given trial values of $\alpha$ and $\beta$, we use the  stellar masses of the 11,419 galaxies of our sample in Equation~\ref{eqn:MHI} to estimate their individual \hi\ masses, ${\MHI}(\alpha,\beta)$. Next, we use these individual \hi\ masses to compute the weighted-average \hi\ mass of the $i$'th subsample, $\langle {\MHI}(\alpha,\beta)\rangle^i$, with the weights being the same as those used to stack the \hii\ emission signals of the subsample. Through this procedure, we effectively obtain the average \hi\ masses of the three subsamples as a function of $\alpha$ and $\beta$, assuming that the $\MHI-\Ms$ relation at $z\approx1$ can be described by  Equation~\ref{eqn:MHI}. The parameters $\alpha$ and $\beta$ are finally obtained by minimising, using a standard steepest-descent approach\footnote{The optimization was carried out using an implementation of the Levenberg-Marquardt algorithm in the scipy package \citep{2020SciPy-NMeth}.}, the $\chi^2$ given by

\begin{equation}
\chi^2 (\alpha,\beta) = \sum_{i=1}^{3} \left( \frac{\langle{\MHI}\rangle^i-\langle {\MHI}(\alpha,\beta)\rangle^i}{\sigma^i_{\MHI}}\right)^2
\label{eqn:chisqr}
\end{equation}

In the above equation, ${\langle\MHI\rangle}^i$ and $\sigma^i_{\MHI}$ are the measurement of the average \hi\ mass in the $i$'th subsample and the uncertainty on the measurement, respectively.

\end{document}